\documentclass[12pt]{article}
\usepackage{amsmath}
\usepackage{graphicx}
\usepackage{enumerate}
\usepackage{natbib}
\usepackage{url} % not crucial - just used below for the URL 
\usepackage{color}

\newcommand{\blind}{1}

\date{August 31, 2023}

\begin{document}

\def\spacingset#1{\renewcommand{\baselinestretch}%
{#1}\small\normalsize} \spacingset{1}

%%%%%%%%%%%%%%%%%%%%%%%%%%%%%%%%%%%%%%%%%%%%%%%%%%%%%%%%%%%%%%%%%%%%%%%%%%%%%%

\if1\blind
{
  \title{\bf Discussion of \\``A Tale of Two Datasets: Representativeness and Generalisability of Inference for Samples of Networks''}
  \author{Nynke M.\ D.\ Niezink\footnote{nniezink@andrew.cmu.edu}\\
    Department of Statistics \& Data Science, Carnegie Mellon University}
  \maketitle
} \fi

\if0\blind
{
  \bigskip
  \bigskip
  \bigskip
  \begin{center}
    {\LARGE\bf Discussion of ``A Tale of Two Datasets: Representativeness and Generalisability of Inference for Samples of Networks''}
\end{center}
  \medskip
} \fi

\spacingset{1.9} % DON'T change the spacing!

\vspace{-1mm}
I congratulate the authors on their timely and insightful article. Since the advent of network analysis, there has been the question of the meaning of sample size in a network setting, which in the context of statistical theory has stirred much academic debate. Traditionally, most applied network studies focused on a single population network -- for example, on the social interactions in one particular tailor shop \citep{kapferer1972strategy} or on the collaboration patterns in one organized crime network \citep[e.g.,][]{campana2018}. More recently, researchers have started collecting populations of networks, with classrooms being the most notable example. In this case, our understanding of asymptotics, inference, and generalizability is more similar to what we are used to in the non-network setting.

Yet, once we have a population of networks, our statistical models may fit some better than others. Also, while techniques like meta-analysis to combine individual networks' estimates or multi-level (hierarchical) modeling work well for a sample of reasonably large networks, they are not easily applicable to smaller networks. The current article addresses some of these challenges, by proposing an Exponential-Family Random Graph Model \citep[ERGM;][]{lusher2013} to jointly model an ensemble of networks, using a multivariate linear model for the ERGM parameters. The authors develop this framework without assuming that all networks in the ensemble are fully observed, which in practice is indeed uncommon. They discuss the requirements for valid inference and present tools for diagnosing a lack of fit in the proposed framework. Network fit is currently often diagnosed by comparing observed but not explicitly modeled network features to the distribution of those features in networks simulated from the estimated model \citep{hunter2008goodness}. However, because of their choice of ERGM parametrization, the authors can leverage existing techniques developed for regression. Apart from elaborating on likely causes and diagnostics for nonidentifiability, they discuss several ways in which a model may fit the data poorly and the corresponding diagnostics.

The article applies the proposed methodology to two household network datasets which were collected in separate surveys. There are two major differences between these surveys. First, in the egocentric ($E$) survey \citep{Hoang2021} only one household member was enrolled (the ego), while in the second survey, the whole household ($H$) was \citep{goeyvaerts2018}. Second, the $H$ survey was restricted to households with a child aged at most 12, but for the $E$ survey, there was no such restriction. The analysis investigates whether or not household members had physical contact over one day given their individual characteristics (age category and gender), household characteristics (e.g., the presence of a child, postal code in Brussels),  and network endogenous effects that are adjusted for network sample size (e.g., triangles).

In this discussion, I take the opportunity to address some potential issues with the modeling and diagnostic framework, focusing on the article's application and the framework's applicability. I hope that some of these observations may lead to further clarification and extensions of the current methodology.

\subsection*{When to ERGM?} Although I generally concur that network data should be analyzed using network methods, with networks of the size analyzed in the article's application, the question arises: to ERGM or not to ERGM? In particular, if we leave out the 2-stars and the triangles effects (and their interactions with the logarithm and the squared logarithm of the network size), the proposed model would reduce significantly to a dyad-independent model -- or edge-independent in this case, as physical contact is an undirected relation. We could obtain maximum likelihood estimates for this trivial ERGM without needing MCMC-based techniques. The focus in the current application is mainly on the effect of household (i.e., network-level) and actor characteristics on the existence of physical contact. As shown in \emph{Model 1d} in the article's Appendix (Table F10), the substantive conclusions on these effects do not change if we leave out the dyad-dependent effects. This is likely related to the fact that more than 28\% of the households comprise only two members. At the same time, I expect the differences in computation time to be significant. In practice, it may therefore be worthwhile to first estimate a dyad-independent model when studying an ensemble of very small networks and only add the dyad-dependent effects in the final model.

The strength of the proposed modeling framework may come to light more when the network endogenous effects are the research focus and the networks studied are a bit larger. For example,  \cite{de2019balance} studied balance theory in the context of sibling-parent-sibling triads. This sociological theory suggests that individuals in triadic configurations prefer to be in a balanced triad, i.e., all relations in the triad are positive or two are positive and one is negative, in line with the idea that `the enemy of the enemy is your friend' \citep{heider1946, heider1958}.  While \cite{de2019balance} focused on triads, the Netherlands Kinship Panel Study  \citep{dykstra2005codebook} their data originates from contains information about larger families (e.g., three generations, new and ex-partners of divorced individuals). Given the social dynamics in family units that experienced divorce, it would be interesting to study balance theory in this larger setting. An extension of the proposed ERGM modeling framework to multiplex networks (in this example, positive and negative ties) would lend itself very well to that.

\subsection*{Network size.} When studying an ensemble of networks, the number of actors per network often varies. Bigger networks usually have lower density (number of ties divided by the potential number of ties), while the networks' average degree (number of ties divided by the number of actors)  is roughly invariant to size. To mimic this behavior, several ERGM parametrizations have been proposed. The article uses the idea of \cite{butts2015} to estimate the effect of network size on density based on the sample of networks, and interacts the linear and quadratic covariates  $\log(n_s)$ and  $\log^2(n_s)$, where $n_s$ is the size of network $s$, with the edges, 2-stars, and triangles effects. 

\begin{figure}
\begin{center}
\includegraphics[width=10cm]{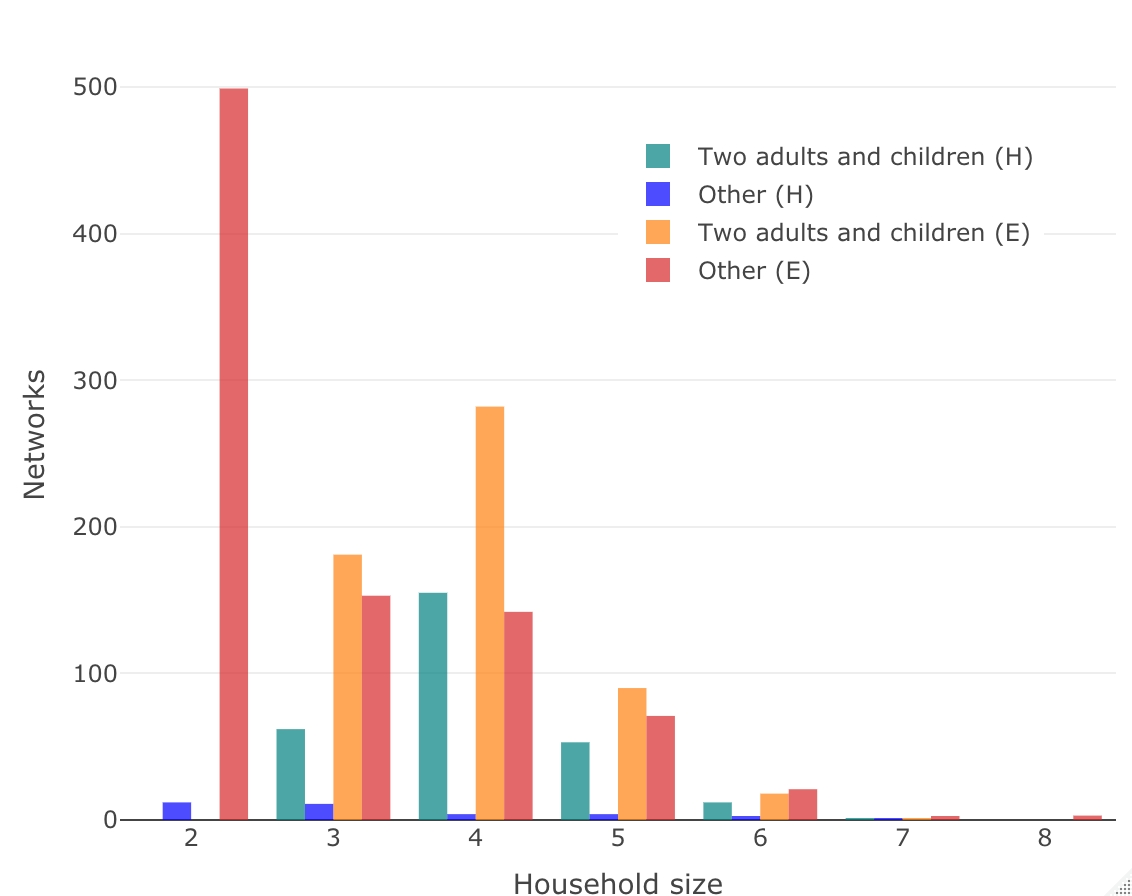}
\end{center}
\caption{Distributions of household sizes in the whole household ($H$) survey and the egocentric ($E$) survey.  In households with two adults and children, there is at least one child.
 \label{fig:HEsizes}}
\end{figure}

While such a parametrization may work well for networks that are fairly homogeneous, such as school classes ranging in size from 20 to 35, I find its applicability conceptually questionable in the current context. Figure~\ref{fig:HEsizes} shows the distribution of household sizes in the egocentric ($E$) and the whole household ($H$) survey. Adults here are defined as individuals older than 18, and children are aged 18 or younger. Note that this constitutes a rough approximation of the role the individuals play within the household: in those households with two individuals over 18 and one or more children, the two adults are likely to be the parents.  Although this approximation does not capture situations such as when adult children are living with their parents, the figure tells an interesting story. In the $E$ dataset, 34.1\% of the households consist of two persons, and in the $H$ data, 48.9\% of the households consist of two adults and two children. Very few households have more than five members. 

Two is a pair, three is a group. That is, the social dynamics that occur among three or more individuals essentially differ from those among pairs of individuals. For networks that are this small, network size could therefore have alternatively been treated as a categorical (e.g., 2, 3, 4, $\geq 5$ individuals) covariate. Moreover, unlike school classes, the networks in the current study are very inhomogeneous, ranging from elderly couples to large families. Future household network analyses should take into account individuals' roles within the household, instead of stratifying by age and gender. If no role information is available, individuals' age gaps could be used as a proxy. 

\subsection*{User guidance.} The authors published their implementation of the model in the R package \verb!ergm.multi!. The availability of this open-source software will be of great help to applied researchers. Nevertheless, the proposed method is not a panacea. It would be good if the authors could comment on when the modeling approach should be preferred over, for example, a hierarchical ERGM \citep{SLAUGHTER2016334} or an ERGM for little networks \cite[ERGM\emph{ito};][]{yon2021}, and when not. Additionally, what should users expect in terms of computation time and how scalable is the methodology?  Finally, the article proposes the use of Pearson residual plots to diagnose model fit and states that, in the household data analysis, these plots indicate a good fit. Yet, there seem to be many outliers, and as the underlying network statistics are small counts close to their exogenous upper bounds, the residuals are skewed downwards and exhibit a striped pattern. This raises the question of what a `bad fit' for an ensemble of small networks would look like. For example, would excluding the network endogenous effects (2-stars, triangles) result in a bad fit? When being introduced to a diagnostic framework, users need to see examples of failure as well as success.

\bibliographystyle{agsm}

\bibliography{Bibliography}
\end{document}